\documentstyle[preprint,aps]{revtex}
\begin{document}
\tightenlines
\title{Anisotropic ``charge-flipping" acceleration of highly charged ions from $(N_2)_n$ clusters in strong optical fields}   
\author{M. Krishnamurthy, D. Mathur, and V. Kumarappan\footnote{Now at: Institute for Physical Sciences and Technology, University of Maryland, Maryland, USA}}
\address{Tata Institute of Fundamental Research, 1 Homi Bhabha Road, Mumbai 400 005, India.}
\date{\today}
\maketitle
\begin{abstract}
The disassembly of molecular clusters $(N_2)_n$ ($n$=50-3000) in strong optical fields is investigated using two-dimensional time-of-flight spectrometry. Very highly charged ions are formed with a two-component energy distribution. A low-energy, isotropic component correlates with Coulomb explosion. A high-energy, anisotropic component, that results from a ``charge flipping" acceleration mechanism, gives rise to ions with energies in excess of the Coulombic limit. 
\end{abstract}
\pacs{36.40.Qv, 52.50.Jm, 34.80.Kw, 36.40.c}

The physics that governs the disassembly of atomic clusters in intense optical fields has attracted considerable attention in the course of the last decade (for a recent review, see \cite{Krainov} and references therein). In the optical regime, light intensities larger than $\sim$10$^{12}$ W cm$^{-2}$ are termed ``intense"; fields generated in such light are strong enough to cause non-perturbative effects in most atoms and molecules. At light intensities above 10$^{14}$ W cm$^{-2}$, multiphoton and/or tunneling effects force nearly every irradiated atom and molecule to undergo ionization. Much effort has been devoted in recent years to studies of small atomic and molecular systems under the influence of such fields. Among the dominant effects observed are above-threshold ionization (ATI), tunneling ionization, high harmonic generation, molecular alignment, and enhanced ionization of molecules. For a recent update on progress in these areas, see Bandrauk {\it et al.} \cite{bandrauk}. Experiments that probe such phenomena are typically conducted under conditions such that the atoms and molecules respond independently to the laser field - the density is kept low enough so that each atom/molecule is ``unaware" of the presence of others in the neighborhood. This is not valid for high-density targets like solids, where macroscopic fields that are generated due to charge separation play a crucial role, and recourse has to be taken to many-body, plasma models in order to gain insight into the dynamics of laser-solid interactions. Gas-phase clusters act as a bridge between the low-density and high-density systems. Large atomic clusters, consisting of several hundred to several hundred thousand atoms per cluster, provide the high density required for substantial absorption of laser energy. Experimentally, it has been discovered over the last few years that when such clusters are exposed to an intense ultrashort laser pulse, they absorb energy very efficiently \cite{boyer}. While energy is absorbed primarily by the electrons in the cluster, it is rapidly redistributed in the form of incoherent radiation (few keV x-rays) and highly energetic ions. The level of ionization and the mean energy of the ejected electrons are both significantly higher than expected from the ionization processes known in isolated atoms and small molecules. The cluster expands and breaks up in a few picoseconds; this rapid disassembly results in the emission of ions with considerable kinetic energy. While maximum ion energies as high as an MeV have been reported \cite{ditmire}, mean energies are typically in the range of 10-100 keV. These are several orders of magnitude higher than the energies that are measured for small molecules. 

Although various theoretical models have been proposed to explain the explosion dynamics of clusters under intense field irradiation, none can adequately explain the experimentally observed features that presently drive research in this area. It is clear that most atoms in a cluster are tunnel ionized at the leading edge of the incident laser pulse. Electrons that are ejected from individual atoms are accelerated away from the cluster by the ponderomotive potential and what gets left behind is a positively charged core that gives rise to an increasing potential barrier to further removal of electrons. The question of whether such a barrier is sufficient to retain a large fraction of electrons or not is a major point of contention between the two major models of laser-cluster interaction. In one of the prevailing scenarios, known in the literature as the hydrodynamic expansion model, it is assumed that retention of most of the electrons by the cluster results in a spherically symmetric, electrically neutral plasma. The retained electrons absorb energy from the laser by collisional inverse bremsstrahlung. The hot electron plasma expands due to hydrodynamic pressure, and transfers energy to the ions.  The alternative scenario is referred to as the Coulomb explosion-ionization ignition model of the dynamics. Here, the electrons that are ejected following tunnel ionization rapidly leave the vicinity of the cluster. As a result, there is a build-up of charge on the cluster that gives rise to a radial field that may become large enough to drive further ionization at the surface of the sphere. The removal of electrons at the surface increases the radial field further and ``ignites" ionization. Cluster explosion due to the Coulombic repulsion between the positively charged ions then occurs. Neither of these scenarios account for several facets of the physics of cluster-laser interactions that have been unveiled in recent experiments. Among them are:  (i) Asymmetric Coulomb explosion of clusters \cite{argon}. (ii) Enhancement in the asymmetry of both the ion and electron emission at the resonance condition, where the plasma frequency matches the driving laser frequency and the energy absorption by the clusters is most effective \cite{xenon}. Recent results have provided the impetus for developing models that extend beyond the prevailing, somewhat simplistic, picture of isotropic, one-dimensional plasma expansion and of probing the possibility that new mechanisms for charged particle acceleration, over and above that due to (a) Coulombic fields in small and moderately-sized clusters and (b) hydrodynamic pressure in large clusters might, contribute to the overall dynamics strong optical fields interact with clusters.  

Apart from fundamental interest from the viewpoint of non-perturbative cluster dynamics, insights into new ion acceleration mechanisms are important as they offer tantalizing scope for applications involving the physics of high energy density matter, and for the development of new charged particle acceleration and ion propulsion schemes. Clusters have also emerged as a promising source of extreme ultraviolet radiation for lithography \cite{euv} and for emission of {\em coherent}  radiation at large multiples of the incident laser frequency \cite{donnelly}.  The interaction of pulses from tabletop lasers with D$_2$ and D-rich clusters has already made possible the study of table-top fusion, which was hitherto restricted to very large national facilities \cite{ditmire2}.  The fusion process also holds promise as a source of mono-energetic and short-pulse neutrons for medical and material science studies \cite{medical}. In order to broaden the scope of studies, we have applied two-dimensional time-of-flight (TOF) spectrometry to probe the disassembly of relatively small {\em molecular} clusters, $(N_2)_n$ ($n$=50-3000), in strong optical fields. These molecular clusters yield very highly charged atomic ions, including hydrogenic N$^{7+}$. The ion energy distribution is found to have two components: a low-energy, isotropic component that correlates with a Coulomb explosion mechanism, and an anisotropic component that results from a ``charge flipping" ion acceleration mechanism that yields ion energies {\em larger} than those expected from purely Coulombic considerations.   

Our experiments were conducted using a supersonic jet expansion source that has been described in earlier reports \cite{argon,xenon,xrays}. Briefly, a piezo-driven solenoid valve, with a 30$^\circ$ conical opening of 500 $\mu$m, was backed with stagnation pressure of up to 14 bar. The central part of the cluster beam was sampled by a 500 $\mu$m skimmer placed $\sim$30 cm downstream, in an interaction chamber maintained at 10$^{-7}$ Torr. A Ti:sapphire laser was used that produces 100 fs long pulses, at a repetition rate of 10 Hz, with a maximum energy of 55 mJ. For the present measurements, 10 mJ pulses were focused to intensities of 10$^{16}$ W cm$^{-2}$ by means of a 25 cm planoconvex lens. A halfwave plate placed just before the lens was used to control the polarization vector of the laser light. Laser-cluster interactions occured $\sim$26 cm downstream of the skimmer. The ions formed when the laser and cluster beams were temporally coincident in the interaction zone were detected by a channel electron multiplier (CEM) placed at the end of a 58 cm long flight tube which was mutually orthogonal to the laser and cluster beams. A three-grid retarding potential energy analyzer (RPA), of length 1 cm, with an intergrid separation of 5 mm, was placed in front of the CEM. Application of appropriate positive voltages on the central grid enabled two-dimensional, charge-discriminated energy spectra to be obtained, in a manner that has been described in earlier reports \cite{argon,water}. Cluster formation was verified by measuring the Rayleigh scattered signal from the clusters using 355 nm light from an Nd:YAG laser. Hagena's scaling law \cite{hagena} was used to estimate cluster size; the average size of $(N_2)_n$ clusters ranged from $n$=50 at a stagnation pressure of 2 bar to $n$=3000 at 10 bar.

Fig. 1 shows the ion energy distribution for two cluster sizes at an intensity of 7$\times$10$^{15}$ W cm$^{-2}$, when the laser light is polarized along the TOF axis as well as perpendicular to it. There is a high-energy component in the distribution (at energies $\geq$3-5 keV) that exhibits a distinct asymmetry: ion energies and fluxes are enhanced when the laser polarization vector is parallel to the TOF axis. The low-energy component of the distribution (at energies $\leq$3-5 keV) is found to be isotropic. This type of behavior is consistent with that observed in atomic clusters \cite{argon,xenon}. However, with $(N_2)_n$ clusters, there is a new feature: a very low-energy component (at energies $\leq$40 eV) that is also asymmetric (see the inset in fig.1). This component has no analog in measurements with atomic clusters and we ascribe it to the ionization signal from unclustered $N_2$ molecules that are present in the supersonic beam. The asymmetry that is observed is attributed to spatial alignment of the molecular axis along the laser polarization vector, a feature of laser-molecule interactions that has been well studied \cite{bandrauk,alignment}. We note that $N_2$ has a very low value of the empirical constant in the Hagena parameter \cite{hagena}, indicating that this molecule has low propensity for clustering;  thus we observe a component of unclustered N$_2$ molecules that also emerges from the skimmer.

Variation of cluster size does not result in alteration of the overall morphology of the energy distribution function but for a shift in the energy value at which the isotropic and anisotropic components meet. This point, which we refer to as the ``knee", shifts progressively towards lower energies as the cluster size is reduced. Fig. 2 shows the experimentally measured ion spectra for different pressures, when the laser light is polarized along the TOF axis.

The maximum value of ion energy, $E_{max}$, as $(N_2)_n$ explodes under Coulombic pressure is from ions that are on the outer surface of the spherical cluster and is given by
\begin{equation}
E_{max} \propto {Q {\bar q}n_c\over{R}},
\end{equation}
where ${\bar q}$ is the average charge per nitrogen ion in a cluster, of radius $R$, comprising $n_c$ nitrogen molecules, and $Q$ is the charge on the ion that is detected. It is established that $n_c$ varies as $P^2$ \cite{xrays,hagena}, where $P$ is the stagnation pressure in the supersonic nozzle and, since $n_c$ scales with cluster size as $R^3$, it is clear that the ion maximum energy, $E_{max}$, should scale as $P^{4/3}$. This analysis is based upon two assumptions: (i) at a given laser intensity, the average charge does not appreciably change with cluster size, and (ii) cluster expansion is such that the radius of the expanding cluster at the end of the laser pulse is proportional to the initial radius. Results of extensive numerical simulations of cluster explosions that Last and Jortner \cite{jortner} have carried out provide ample justification for the validity of both assumptions. Fig. 3 depicts the ``knee" energy that we have measured for different values of stagnation pressure, $P$. A fit to our data (solid line in fig. 2) shows that the ``knee" energy scales as $P^\alpha$, where $\alpha$=1.5$\pm$0.2, which is in consonance with expectations from the our analysis. This brings to the fore the necessity of having to invoke an additional acceleration mechanism, over and above Coulombic pressure, to account for ions that possess energies in excess of the ``knee" energy. By working with clusters of relatively small size we exclude the possibility of hydrodynamic pressure coming into play. In this connection it has not escaped our attention that ion energies at values less than the knee energy scale approximately quadratically with ion charge for the clusters under study. 

The anisotropy that is observed for the highest energy ions, beyond the ``knee" energy, provides a clue as to the nature of this additional acceleration mechanism. As has been observed in earlier work on $Ar_n$ clusters \cite{argon}, and in numerical simulations of explosions of clusters such as $Xe_{147}$ \cite{blenski}, that the charge state of an ion can rapidly change in resonance with the applied optical field. The radial field due to the charged cluster and the applied laser field act in the same direction for half the optical cycle, and a high ionic charge state can be attained. But, in the next half cycle, these two fields oppose each other, and the net charge is reduced from its maximum value due to ion-electron recombination. This charge flipping gives rise to a net cycle-averaged force that serves to accelerate the ions. The simulations carried out by Ishikawa and Blenski for $Xe_{147}$ \cite{blenski} showed that such acceleration is much smaller for ions that are ejected perpendicular to the laser polarization vector. Measurements carried out by us \cite{argon} on $Ar_{40000}$ demonstrated asymmetry in the high energy component of $Ar^{q+}$ ($q\leq$8) ions produced in the Coulomb explosion regime, and subsequent measurements on explosions of $Xe_{150000}$ in the hydrodynamics expansion regime indicated that the asymmetry is likely to be a feature that is independent of the regime in which the explosion takes place. The present measurements on small $(N_2)_n$ clusters provide verification of the earlier conjectures. 

In summary, measurements of  ion energies of the explosion products from $(N_2)_n$ ($n$=50-3000) clusters provide evidence for formation of energetic ions in very high charge states. The ion energy distribution has a low-energy isotropic component, whose maximum energy is accounted for by simple Coulombic considerations. Over and above this, there is a not-insubstantial flux of ions that have energy in excess of the Coulombic limit. Our measurements provide evidence for a  charge-flipping mechanism that accounts for the additional acceleration that is experienced by the exploding ions. It also rationalizes the asymmetry with respect to laser polarization that is observed for these high energy ions. Exploitation of this additional acceleration mechanism might be of utility in future considerations of tabletop accelerators.

\begin{figure}
\caption{Ion energy spectra for laser polarizations parallel (0$^{\circ}$) and perpendicular (90$^{\circ}$) to the TOF axis for $(N_2)_{2300}$ and $(N_2)_{1200}$ clusters at a laser intensity of $7{\times}10^{15}$ W cm$^{-2}$. Note that for energies less than $\sim$4 keV, the two spectra overlap ; differences are only seen for high-energy components.  Note that the ``knee" energy shifts from 10 keV in the case of the larger cluster to 4 keV for the smaller one.The inset shows the asymmetry at very low energies ($\leq$40 eV). This features is ascribed to spatial alignment of unclustered $N_2$ molecules.}
\end{figure}

\begin{figure}
\caption{ Ion energy spectra for laser polarization parallel to the TOF axis for different values of stagnation pressures, $P$. The ``knee" energy becomes smaller as the stagnation pressure is lowered.}
\end{figure}  

\begin{figure}
\caption{Variation of ``knee" energy with stagnation pressure, $P$. The solid line is a fit to the measured data and indicates a $P^\alpha$ ($\alpha$=1.5$\pm$0.2) dependence. Cluster sizes for few vales of $P$ are indicated by the vertical arrows.} 
\end{figure}

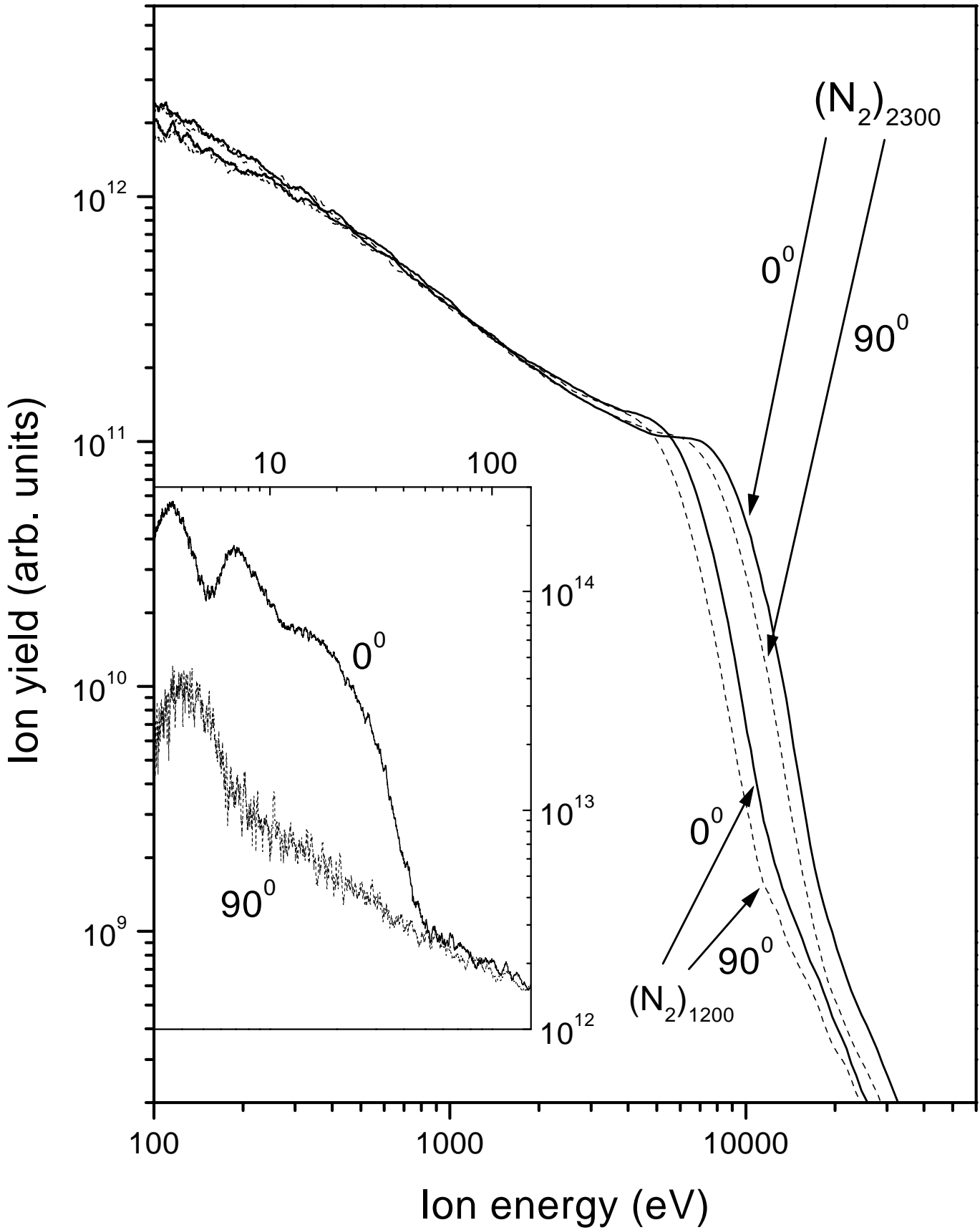
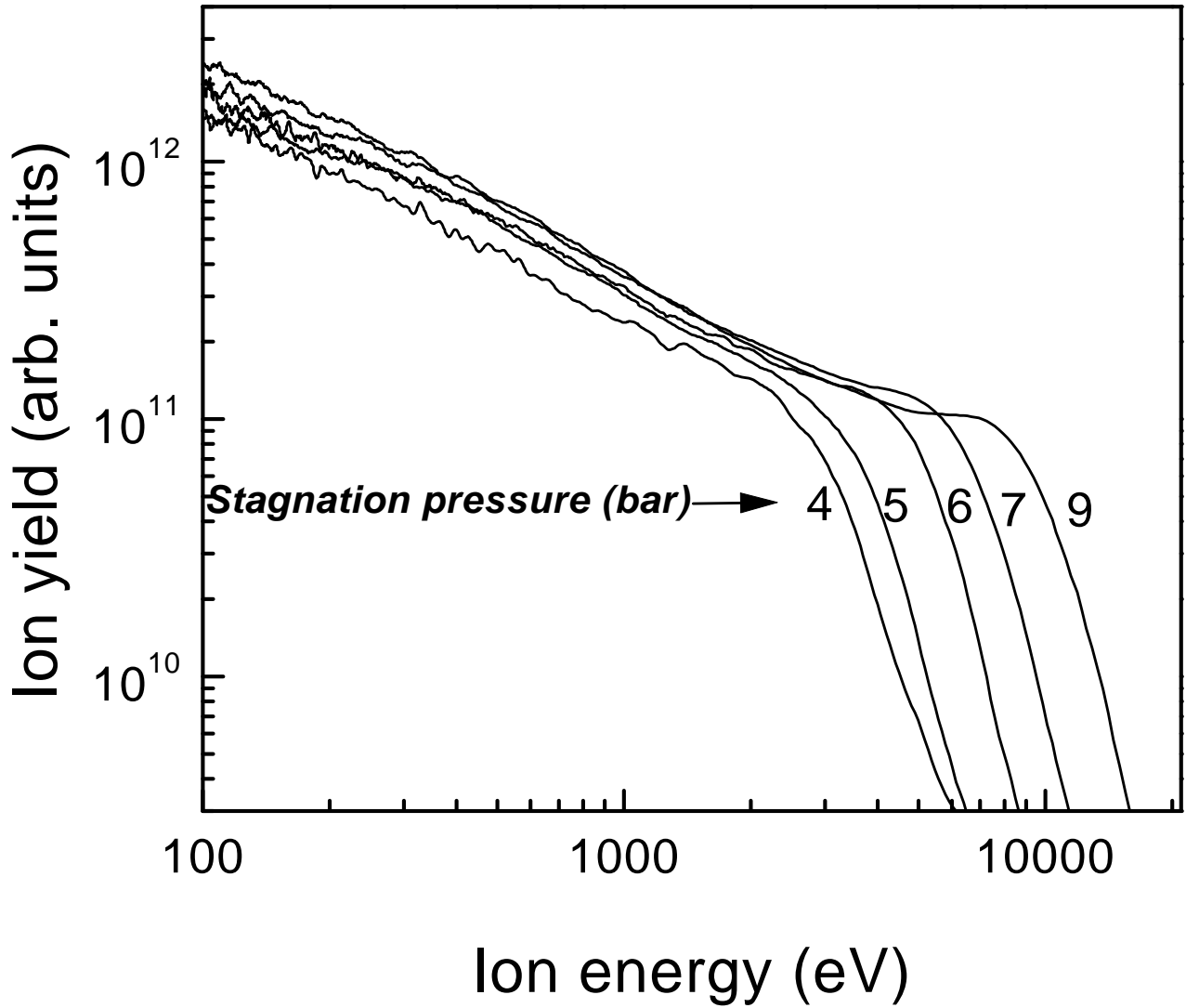
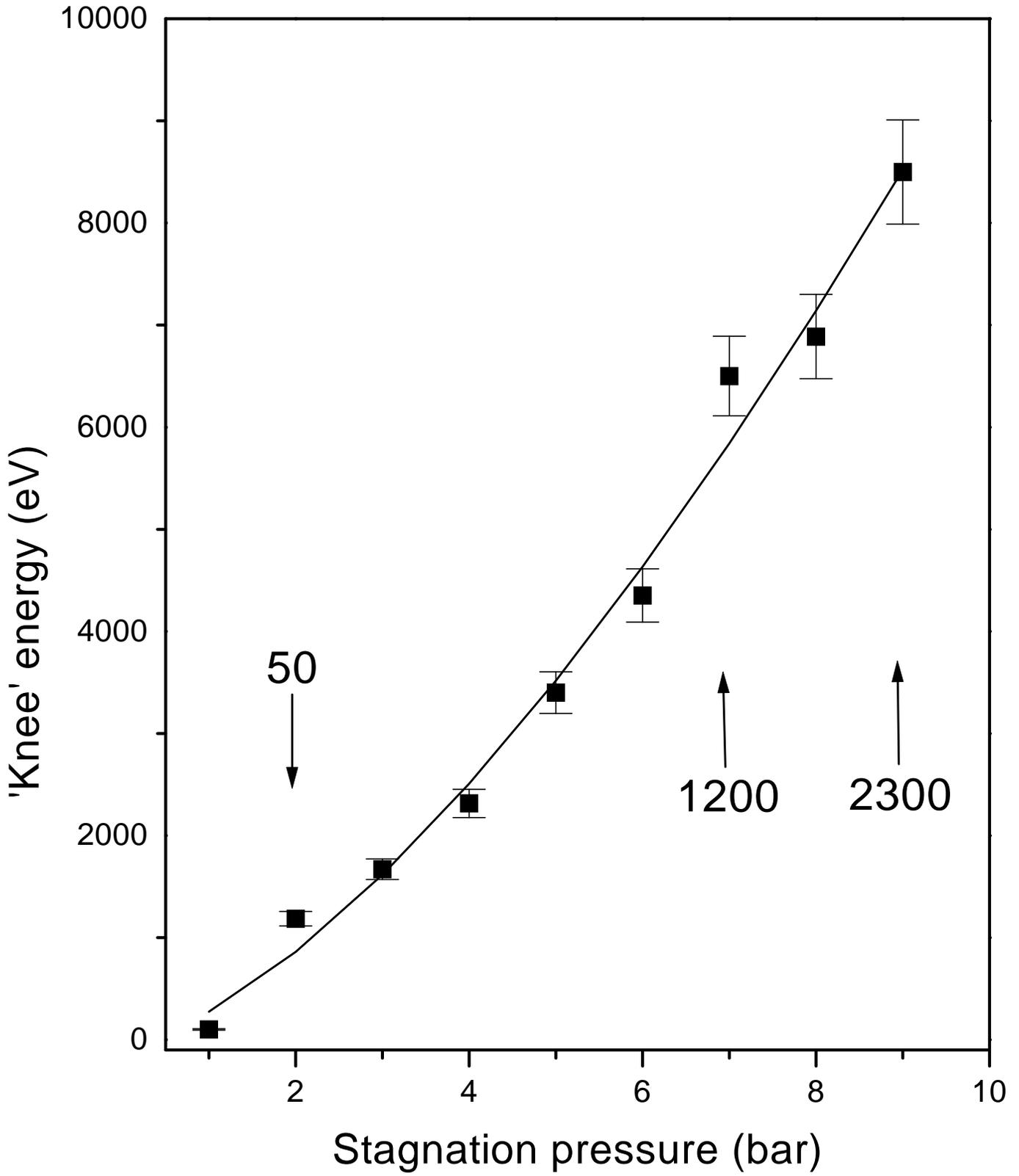
\end{document}